\documentclass[osajnl,twocolumn,showpacs,superscriptaddress,10pt]{revtex4-1} 
\usepackage{amsmath,amssymb,graphicx}

\begin{document}

\newcommand{\be}{\begin{equation}}
\newcommand{\ee}{\end{equation}}
\newcommand{\C}{\mathcal}
\newcommand{\R}{\mathrm}

\newcommand{\eref}[1]{(\ref{#1})}
\newcommand{\sref}[1]{section~\ref{#1}}
\newcommand{\fref}[1]{figure~\ref{#1}}
\newcommand{\tref}[1]{table~\ref{#1}}
\newcommand{\Eref}[1]{Equation (\ref{#1})}
\newcommand{\Sref}[1]{Section~\ref{#1}}
\newcommand{\Fref}[1]{Figure~\ref{#1}}
\newcommand{\Tref}[1]{Table~\ref{#1}}

\title{Simultaneous Unbalanced Shared Local Oscillator Heterodyne Interferometry (SUSHI) for high SNR, minimally destructive dispersive detection of time-dependent atomic spins}

\author{Mary Locke}
\author{Chad Fertig}\email{Corresponding author: chad.fertig@honeywell.com}
\affiliation{Department of Physics and Astronomy, University of Georgia, Athens, Georgia 30602}

\begin{abstract}
We demonstrate ``Simultaneous Unbalanced Shared Local Oscillator Heterodyne Interferometry (SUSHI),'' a new method for minimally destructive, high SNR dispersive detection of atomic spins.  In SUSHI a dual-frequency probe laser interacts with atoms in one arm of a Mach-Zehnder interferometer, then beats against a bright local oscillator beam traversing the other arm, resulting in two simultaneous, independent heterodyne measurements of the atom-induced phase shift.  Measurement noise due to mechanical disturbances of beam paths is strongly rejected by the technique of \emph{active subtraction} in which anti-noise is actively written onto the local oscillator beam via an optical phase-locked-loop.  In SUSHI, technical noise due to phase, amplitude, and frequency fluctuations of the various laser fields is strongly rejected (i) for any mean phase bias between the interferometer arms, (ii) without the use of piezo actuated mirrors, and (iii) without signal balancing.    We experimentally demonstrate an ultra-low technical noise limited sensitivity of 51~nrad$/\sqrt{\R{Hz}}$ over a measurement bandwidth of 60~Hz to 8~kHz using a 230 $\mu$W probe, and stay within $\sim$3 dB of the standard quantum limit as probe power is reduced by more than 5 orders of magnitude to as low as 650~pW.  SUSHI is therefore well suited to performing QND measurements for preparing spin squeezed states and for high SNR, truly continuous observations of ground-state Rabi flopping in cold atom ensembles.
\end{abstract}

\maketitle 

\section{Introduction: Quantum limited dispersive measurements of collective atomic spins}
\label{sec:intro}

In dispersive detection of trapped atomic gases, information on the quantum state of the atoms is gained via a measurement of the phase shift induced on a detuned probe laser passing through the cloud \cite{Hope:2005kx}.  Dispersive detection has been used to observe spinor dynamics in Bose-Einstein condenses \cite{Andrews:1996dq, Ketterle:1999hc, Liu:2009vo}, to perform continuous observations of coherent Rabi flopping of atomic spins \cite{Windpassinger:2008cr, Windpassinger:2008tg}, and to prepare spin-squeezed states via quantum non-demolition (QND) measurements \cite{Kuzmich:1999kl, Oblak:2005nx, Appel:2009oq}.  Continuous, weak dispersive measurements of atomic spins may also allow for higher stability operation of microwave atomic clocks \cite{Shiga:2012uq}.

The signal-to-noise ratio (SNR) for dispersive detection depends on many experimental parameters.  In general, the SNR of any interferometric measurement of the phase shift induced on a coherent state probe laser can be expressed in the form \cite{Chan:1983de, Shapiro:1984jp, Shapiro:1985gd, Leong:1986wb}
\be
\label{eq:SNRP}
\R{SNR}\sqrt{B} = K\Delta \phi \sqrt{2\kappa}\sqrt{\C{P}_{probe}}
\ee
where $B$ the measurement bandwidth, $\Delta \phi$  is the mean detected phase shift, $\kappa$ is the quantum efficiency of the photodetector, $\C{P}_{probe}$ is the probe laser power in units of photons/s at the probe wavelength ($\C{P}=P/\hbar\omega$), and $K$ is a factor which depends on probe power, detector noise and other experimental details.   When the quantum uncertainty of the probe laser field domates the measurement uncertainty, the SNR takes its maximal ``standard quantum limited'' (SQL) value.   For heterodyne phase detection this corresponds to $K$ taking the limiting value $K_{het,ideal}=1$, while for homodyne detection it corresponds to $K_{hom,ideal}=\sqrt{2}$ \cite{Shapiro:1985gd, Haus:1995el}.   The scaling of SNR with $\sqrt{\C{P}_{probe}}$ reflects a general number-phase uncertainty principle for measurements on coherent state fields \cite{Dariano:1994hu, Ou:1996jg, Luis:1996bb, Ou:1997ks, Opatrny:1998iu}.  In turn, this scaling implies that the SQL SNR scales as the square-root of the spontaneous scattering rate, a situation sometimes referred to as ``minimally destructive''\cite{Lye:2003vl}.  In any real experiment, excess technical noise will inevitably depress $K$ (possibly far) below the SQL limiting value.  Furthermore, any given scheme is apt to achieve SQL SNR only over a limited range of probe powers and/or detunings, and therefore may not be suitable for all applications.  As an example, the continuous weak observation of coherent driven Rabi flopping between hyperfine states may require very low scattering rates ($<$1ph/atom/sec) and probe detunings less than 100$\Gamma$.  In contrast, a higher SQL SNR is required for the (stronger) QND measurements needed to prepare spin squeezed states, thus demanding the use of brighter probes.

It is useful to categorize the various (non-imaging) dispersive detection methods reported in the literature as either ``1-arm'' or ``2-arm''.  In 1-arm schemes the total optical power falling on the final photodetector also passes through the atom cloud.  Such schemes either measure the shift in the relative phase of two orthogonal polarizations \cite{Kuzmich:1999kl, Smith:2003ue} or two different frequencies ~\cite{Lye:2004ys, Kohnen:2011} simultaneously present in the probe beam.  As shown in Appendix \ref{sec:1vs2}, 1-arm schemes must employ relatively bright probes to overcome detector dark noise, and therefore typically exhibit  $\sim$kHz spontaneous scattering rates.  In 2-arm schemes, one of the two interfering optical fields bypasses the atom cloud altogether and is recombined with the other (the "probe") just before the final final photodetector.  By making the by-passing field bright, 2-arm schemes are capable of approaching SQL SNR for probes fields as dim as 3 pW \cite{Figl:2006}.  However, 2-arm schemes are susceptible to path noise --- mechanical disturbances of the interferometer optics.  One method of suppressing path noise is described in Refs. \cite{Windpassinger:2008cr, Windpassinger:2008tg, Figl:2006}, in which beam paths are actively stabilized using a servo-controlled mirror and an auxiliary, far detuned laser.  A different solution is implemented by Appel et al.~\cite{Appel:2009oq}, wherein path noise is passively rejected by subtracting the (baseband) output of two spatially coincident homodyne interferometers made with a dual-frequency probe beam.  In this scheme, path noise induces identical phase shifts to each probe component, while the atoms induce oppositely signed phase shifts of equal magnitude.  By differencing the two interferometer signals, path noise is rejected and the atom signal revealed.   However, strong path noise rejection requires careful matching of the two interferometer signals, and good rejection of laser intensity noise and frequency noise requires the interferometer to be operated precisely at the white light position.  It is impossible to maintain these conditions if the phase shift to be measured is \emph{itself} large and time-varying.

To enable continuous, quantum limited, dispersive measurement of dynamically evolving atomic spins over a wide range of probe powers, we have developed ``Simultaneous Unbalanced Shared local oscillator Heterodyne Interferometry'' (SUSHI).   In SUSHI, two simultaneous and overlapping heterodyne interferometers are made between a dual-frequency probe beam and a (bright) local oscillator, which bypasses the atoms.  In this regard, SUSHI resembles the scheme of Ref.~\cite{Appel:2009oq}.  However, in SUSHI, the atom-induced phase shift of the probe light is revealed, and path noise, laser intensity noise and laser frequency noise rejected, by a technique we call \emph{active subtraction}, without the use of auxiliary far-detuned lasers, active mirrors, precision balancing of the detector responses, nor any stipulation that the white light condition must be met in any interferometer.
Moreover, SUSHI can maintain its ultra-low technical phase noise floor even in the presence of large, time-varying atom-induced phase shift signals.  Thus it is ideally suited to performing real-time, high SNR quantum weak measurements of time-evolving atomic magnetizations.

\section{Simultaneous Unbalanced Shared local oscillator Heterodyne Interferometry (SUSHI)}
\label{sec:SUSHIintro}
\subsection{General Overview}

In SUSHI, a single spatial mode probe laser beam, composed of two frequency components (referred to as the upper- and lower-sidebands---USB and LSB), passes through an atom cloud (see \Fref{fig:opticalcircuit}).  The sidebands bracket an optical resonance (not necessarily symmetrically) and accumulate phase shifts of opposite sign due to the atomic interaction, similar to Ref.~\cite{Appel:2009oq}.  (We comment here that while the condition of oppositely phase shifted sidebands does not underpin the mechanism by which technical noise is rejected by SUSHI, the standard quantum limited SNR of \eqref{eq:SNR(SUSHI,sql)} cannot be achieved without it.)

The relative phase of the sidebands is pre-stabilized (i.e, \emph{before} interacting with the atoms) by an optical phase locked loop (OPLL).   After passing through the atoms the probe beam is combined with a bright local oscillator (LO) beam on a fast photoreceiver.  The LO bypasses the atoms, and so does not disturb the cloud.  The RF  photocurrents of the USB$\otimes$LO and LSB$\otimes$LO optical beatnotes are separated by filtering, and their phases are measured relative to a master reference synthesizer.  In this way two independent heterodyne measurements of the atomic dispersion are made simultaneously.  To achieve strong rejection of relative path noise between the two arms, we employ active subtraction, in which a second OPLL locks the phase of the LO field to the phase of the USB field \emph{after} the USB field has passed through, and been phase shifted by, the atom cloud. Mechanical disturbances do not introduce relative phase noise between the LSB and USB fields because they occupy the same spatial mode over their entire path through the atoms.  The effect of the active subtraction is to endow the output of the 2-arm LSB$\otimes$LO interferometer signal with the intrinsic vibration insensitivity of a 1-arm measurement, while revealing the total differential atom-induced phase shift between the probe beam's two spectral components with quantum limited sensitivity.

\subsection{The SUSHI Signal}

\begin{figure*}
\includegraphics*[width=6.7 in]{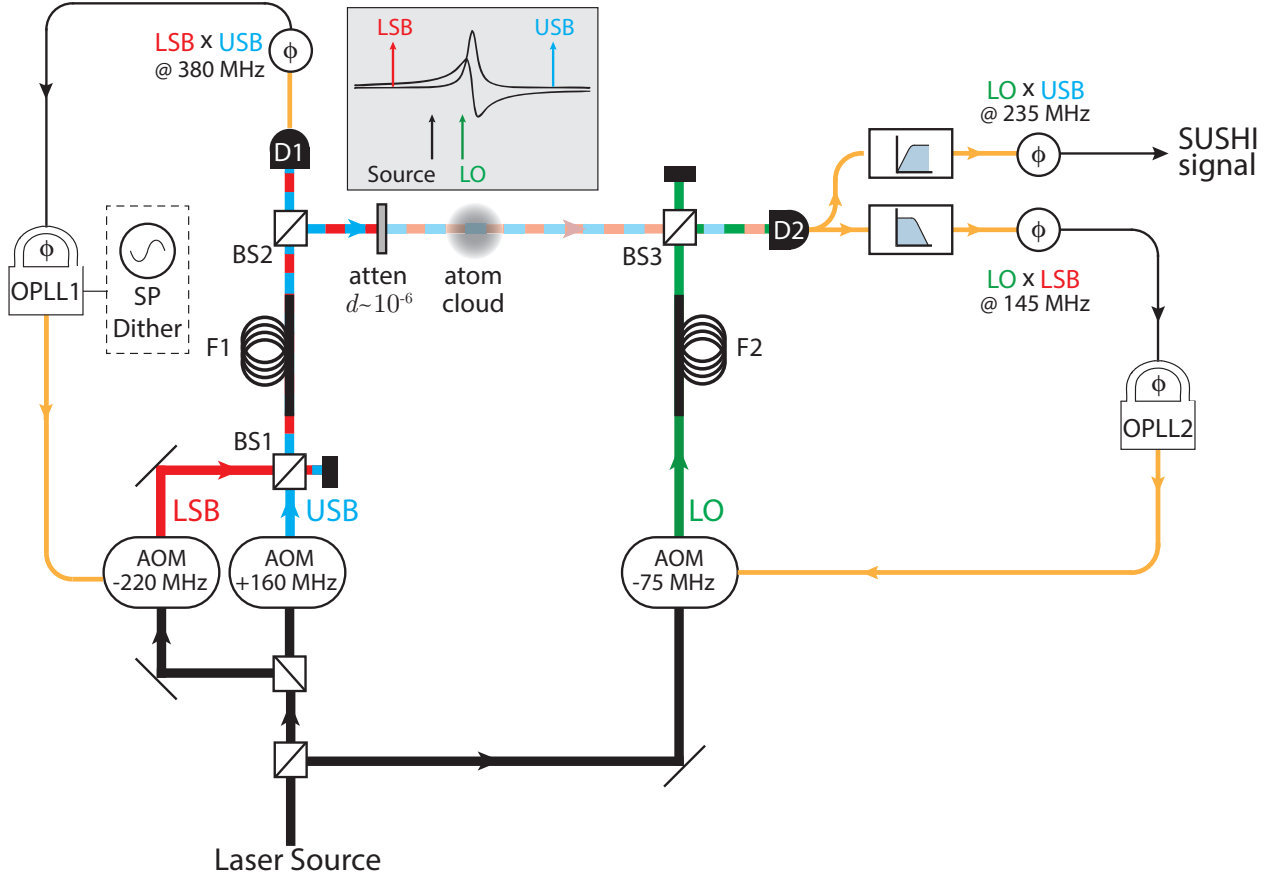}
\caption{\label{fig:opticalcircuit} Simplified schematic of SUSHI detection.  Thick lines represent optical paths (in fiber and/or free space), orange lines represent RF frequency signal paths, and thin light black lines represent baseband signal paths.  Padlock icons represent the electronic signal processing for optical-phase-locked loops (OPLLs).  Inset depicts detunings (not to scale) of the laser source and LSB, USB and LO optical fields relative to an isolated atomic optical resonance.  The dashed box denotes the set-point dither circuit used in bench-test measurements of the SNR for SUSHI; it can introduce a time-dependent phase modulation between the two probe beam colors, simulating the effect on the probe sidebands of an ensemble of atomic spins undergoing driven Rabi-flopping.}
\end{figure*}

In our implementation of SUSHI all three required optical fields are derived from a single laser source (see \Fref{fig:opticalcircuit}).  First, the two beams destined to become the LSB and USB fields are picked-off from the source beam, frequency shifted in separate acousto-optic modulators (AOMs), recombined on beamsplitter BS1, and launched into a single mode optical fiber, forming the probe beam.   This procedure inevitably introduces relative phase noise between the sidebands; to re-establish phase coherence an optical-phase-locked loop (OPLL1) is employed, in which the LSB AOM  is used as a phase actuator to actively steer the phase of LSB to follow that of USB, up to a small residual phase noise $\epsilon_\R{rpn}$.  In our system this residual noise has approximately equal contributions from the RF frequency photodetection dark noise and photon shot noise at D1, and baseband frequency residual laser amplitude noise.  Up to constant phases (which do not affect heterodyne detection) we have for the probe fields at D1

\begin{subequations}
\begin{align}
\R{LSB^{(D1)}} &= t_{2} E_\R{lsb} \cos\left[\omega_\R{lsb} t + \epsilon_\R{rpn}\right]
\label{eq:lockedfirst}\\
\R{USB^{(D1)}} &= t_{2}E_\R{usb} \cos\left[\omega_\R{usb} t\right]
\label{eq:lockedsecond}
\end{align}
\end{subequations}
where we choose as our (instantaneous) global phase zero the USB field at D1, to which we will refer any time-dependent phase shifts (i.e., noise or signal) experienced by any other fields.

Beamsplitter BS2 reflects $(r_{2})^{2}=1-({t_{2})^{2}}=4$\% of the probe beam power towards the atoms.  Prior to passing through the atoms cloud the probe beam is attenuated by an additional factor $d\approx10^{-6}$.  After passing through the atom cloud the probe beam falls on the final beamsplitter BS3, which transmits $(t_{3})^{2}=96$\% of its power to photodetector D2. At this point the two probe sideband fields have acquired a \emph{common} phase noise  $\epsilon_\R{sh}$ due to path noise (``shaking'').  They are also imprinted with (potentially time-dependent) phase shifts induced by the atom cloud.  Anticipating the results of \Sref{sec:SUSHIwith2levelgasses}, we remark here that for any fixed achievable sideband splitting, and for a two-level atom, it is easy to show, using Equations \eref{eq:SUSHI2L} and \eref{eq:SUSHISLGamma}, that the SUSHI signal per scattered photon is optimized for symmetric detuning about the atomic resonance, yielding atom-induced phase shifts of equal magnitude and opposite sign on each sideband. These assumptions also conveniently simplify many expressions in this section, and so we make them here.
\be
\Delta \phi_\R{usb}(t) = -\Delta \phi_\R{lsb}(t) \equiv \Delta \phi_\R{cl}(t)
\label{eq:deltacloud}
\ee
When implementing SUSHI detection of multi-level atoms (e.g., \Sref{sec:m}), no general statement can be made about the optimality of symmetric sideband detuning, nor about the condition of equal magnitude atom-induced phase shifts; however, the condition of opposite phase shifts is universally optimal for detecting any phase object using SUSHI.

We assume the phase shift induced by the atomic cloud satisfies $\Delta \phi_\R{cl}\leq\pi$, which is the natural operating regime for dispersive detection.  We therefore have for the probe fields at D2,
\begin{subequations}
\begin{align}
\R{LSB^{(D2)}} &= d r_{2}t_{3} E_\R{lsb} \cos\left[\omega_\R{lsb} t + \epsilon_\R{sh}  - \Delta \phi_\R{cl}(t)  + \epsilon_\R{rpn}\right]\label{eq:D2first}\\
\R{USB^{(D2)}} &= d r_{2}t_{3} E_\R{usb} \cos\left[\omega_\R{usb} t + \epsilon_\R{sh} + \Delta \phi_\R{cl}(t)\right]\label{eq:D2second}
\end{align}
\end{subequations}
The $\R{LO}$ beam follows a different path to D2, over fibers and free space, bypassing the atoms, and inevitably acquires a phase noise  $\epsilon_\R{lo}$ relative to probe beam.  BS3 reflects  $(r_{3})^{2}=1-({t_{3})^{2}}=4$\% of the LO power to D2:
\be
\R{LO}^{(D2)}  = r_{3}E_\R{lo} \cos\left[ \omega_\R{lo} t + \epsilon_\R{lo} \right]
\label{eq:LO}
\ee
All three fields---USB, LSB and LO---overlap and mutually interfere on photodetector D2, generating a photocurrent with 3 DC terms and 3 RF terms:
\be 
I_{\R{tot}}  = \rho \left(\R{LO^{(D2)}} +\R{LSB^{(D2)}} + \R{USB^{(D2)}}\right)^2 = I_{\R{RF}} + I_{\R{DC}},
\ee
where $\rho$ is the photodetector response (dimensions of current$/E^{2}$).  We focus on the RF terms, which we write as
\be
\label{eq:Irf}
I_{\R{RF}} = I_{\R{lo \otimes lsb}} + I_{\R{lo \otimes usb}} + I_{\R{lsb \otimes usb}}
\ee
where
\begin{subequations}
\begin{align}
\label{eq:IrfB}
I_{\R{lo \otimes lsb}} =  I^{0}_\R{lo \otimes lsb}\cos [&- \epsilon_\R{lo}+ \epsilon_\R{sh} +\epsilon_\R{rpn}  - \Delta \phi_\R{cl} \nonumber \\ 
&- t(-\omega_\R{lo}+\omega_\R{lsb})]\\
\label{eq:IrfC}
I_{\R{lo \otimes usb}} =  I^{0}_\R{lo \otimes usb}\cos [&- \epsilon_\R{lo} +\epsilon_\R{sh} + \Delta \phi_\R{cl}\nonumber\\ 
 &- t (-\omega_\R{lo}+\omega_\R{usb})]
\end{align}
\end{subequations}
In the above equations, $I^{0}_\R{lo \otimes lsb}=d r_{2}t_{3} r_{3}\rho E_\R{lo}  E_\R{lsb}$ and $I^{0}_\R{lo \otimes usb}=d r_{2}t_{3}r_{3} \rho E_\R{lo}  E_\R{usb}.$  The weak beatnote between the two dim sidebands, $I_{\R{lsb \otimes usb}}$, is hereafter ignored.

It is essential to SUSHI that the LO optical frequency not be centered between the sidebands: $\omega_\R{lo} \neq \frac{1}{2} (\omega_\R{lsb}-\omega_\R{usb})$, so that the three RF photocurrents of \eref{eq:Irf} are at different frequencies and can be separated by electronic filters.  We measure the phases of the photocurrents relative to a master RF reference in separate digital phase-frequency detector (PFD) circuits.  (Digital phase detection is (1) independent of signal power over a range of 30dB, (2) strongly rejects technical amplitude noise, regardless of the phase operating point, and (3) has a large and nearly constant gain factor $m_\R{pfd}=dV_\R{pfd}/d\phi$ over a user-selectable range.)  Collecting common phase noise terms into a single term $\epsilon_\R{com}\equiv - \epsilon_\R{lo} + \epsilon_\R{sh}$, we have at the outputs of the two separate PFD circuits:
\begin{subequations}
\label{eq:Vpfd}
\begin{align}
V_\R{pfd}(\phi_{\R{lo} \otimes \R{lsb}}) = m_\R{lsb}  \bigl[&\epsilon_\R{com} - \Delta \phi_\R{cl}(t) + \epsilon_\R{rpn}\nonumber \\
& + \epsilon_\R{het}\left(I^{0}_\R{lo\otimes lsb}\right)\bigr]  \label{eq:Vlsb}\\
V_\R{pfd}(\phi_{\R{lo} \otimes \R{usb}}) = m_\R{usb} \bigl[&\epsilon_\R{com} + \Delta \phi_\R{cl}(t)  \nonumber \\
&+ \epsilon_\R{het}\left(I^{0}_\R{lo\otimes usb}\right)\bigr] \label{eq:Vusb}
\end{align}
\end{subequations}
where $m_\R{xsb}\epsilon_\R{het}(I^{0}_\R{lo\otimes xsb})$ is the total noise, referred to the output of the phase detector, of the heterodyne phase detection of the XSB$\otimes$LO beatnote (X=U,L), including photon shot noise, detector dark noise, and any excess electronic noise.

If the phase detector gains were perfectly matched (i.e., $m_\R{lsb}= m_\R{usb}$), one could generate the final SUSHI signal, and achieve complete cancellation of $\epsilon_\R{com}$, by direct subtraction of the two baseband signals \eref{eq:Vlsb} and \eref{eq:Vusb}:
\begin{align}
V_\R{SUSHI}&= V_\R{pfd}(\phi_{\R{lo} \otimes \R{lsb}})-V_\R{pfd}(\phi_{\R{lo} \otimes \R{usb}})\nonumber\\
 &= m [2 \Delta \phi_\R{cl} + \epsilon_\R{rpn} +\epsilon_\R{het}(I^{0}_\R{lo\otimes lsb})+\epsilon_\R{het}(I^{0}_\R{lo\otimes usb})]\nonumber\\
 &\equiv\C{V}_\R{SUSHI} + \epsilon_\R{SUSHI}
 \label{eq:directsubtract}
\end{align}
where the mean SUSHI signal $\C{V}_\R{SUSHI}=2 m \Delta \phi_\R{cl}$ is seen to be directly proportional to the total differential atom-induced phase shift between the sidebands,  but is independent of their mean amplitudes, and the SUSHI noise $\epsilon_\R{SUSHI}= m[\epsilon_\R{rpn} + \epsilon_\R{het}(I^{0}_\R{lo\otimes lsb})+\epsilon_\R{het}(I^{0}_\R{lo\otimes usb})]$ is free of $\epsilon_\R{com}$.  However it is difficult to maintain the balanced gains condition to better than 1\%, which does not provide sufficient rejection to achieve quantum limited performance without taking additional measures to actively stabilize the interferometer arms.  Moreover, even with perfect signal balancing, direct subtraction in the manner of \eref{eq:directsubtract} can still fail if the path noise fluctuations are larger than the dynamic range of the digital phase detection circuit.  

To solve these problems and achieve strong rejection path noise from the SUSHI signal, we have developed a scheme of \emph{active subtraction}, wherein the $V_\R{pfd}(\phi_{\R{lo} \otimes \R{lsb}})$  signal is used as the error signal of a second optical phase locked loop (OPLL2) which writes the phase deviation $\delta \phi=\epsilon_\R{com}-\Delta \phi_\R{cl}(t)+\epsilon_\R{het}(I^{0}_\R{lo\otimes lsb})$ directly onto the optical phase of the LO field.  Note that this deviation contains both signal and (anti-)noise.  With OPLL2 locked, the LO field at detector D2 becomes (cf. Eq.~\eref{eq:LO})
\begin{align}
\R{LO}^\R{(D2,locked)}=  r_{3}E_\R{lo} \cos [ \omega_\R{lo} t &+\epsilon_\R{sh} -\Delta \phi_\R{cl}(t)  \nonumber \\
&+\epsilon_\R{het}(I^{0}_\R{lo\otimes lsb})],
\end{align}
and the final SUSHI signal is just the output of the $\R{LO}\otimes\R{USB}$ phase detector directly:
\begin{align}
&V_\R{SUSHI}=V_\R{pfd}(\phi_{\R{lo} \otimes \R{usb}})   \nonumber \\
&=m_\R{usb} [2 \Delta \phi_\R{cl}+\epsilon_\R{rpn} +\epsilon_\R{het}(I^{0}_\R{lo\otimes lsb})+\epsilon_\R{het}(I^{0}_\R{lo\otimes usb})],
\label{eq:Vsushi}
\end{align}
which is the same as \eref{eq:directsubtract}.  We analyze the three surviving noise terms in $\epsilon_\R{SUSHI}$ in the next section.  We do not here include the effects of constant or slowing driftingphase offset of the PFD circuits, as we intend SUSHI for dynamical measurements in which such offsets and drifts would naturaly be removed in the data post-processing. If absolute phase differences are desired, proper normalization in the form of an ``atom minus no-atom'' data taking would be required. This consideration is common to all interferometric phase measurements of atom clouds, and not peculiar to SUSHI.

Finally, we note that the SUSHI signal generated through active subtraction exhibits the same insensitivity to master synthesizer phase fluctuations generally manifested by any dual-mixer, phase-differencing measurement.

\subsection{SNR of SUSHI with noisy detectors}
\label{sec:detlimits}
In this section we derive the sensitivity limit for measuring small atom-cloud induced phase shifts using SUSHI, taking into account all major  sources of fundamental and technical noise.  The SNR of the SUSHI signal \eref{eq:Vsushi} will be the same as that of \eref{eq:directsubtract} if OPLL2 is shot noise limited.  \Eref{eq:directsubtract}, in turn, is the difference of two independent heterodyne measurements of the atom cloud, each made with one of the sideband fields.  These two separate heterodyne measurements, at different RF frequencies, have uncorrelated noise.   The SNR of a single heterodyne phase measurement of $\Delta\phi_\R{cl}$ may be calculated as the ratio of the mean heterodyne RF photocurrent to the uncorrelated sum of all RF noise currents (see \Eref{eq:SNR1and2}).  Therefore, under the conditions of \Eref{eq:deltacloud}, we find that the SUSHI phase shift signal is twice the phase shift experienced by one sideband, while the noise is $\sqrt{2}$ larger (i.e, their noise powers add).  Thus, the SNR for a SUSHI measurement of the atom cloud is given by:
\begin{align}
\label{eq:SNR}
&\R{SNR(V_\R{SUSHI})} = \nonumber\\
&\frac{2}{\sqrt{2}} \times\frac{\Delta\phi_\R{cl}\kappa\sqrt{2\C{P}_\R{lo}}\sqrt{2\C{P}_\R{xsb}}}
{\sqrt{B}\sqrt{2\kappa\C{P}_\R{lo} + 2\kappa\C{P}_\R{xsb} + \C{S}_{I^{2}\R{det}}+\C{S}_{I^{2}\R{rpn}}}}.
\end{align}
Here, $\C{S}_{I^{2}\R{det}}$ is the  ``current-squared'' spectral density of dark noise in detector D2,  $\C{S}_{I^{2}\R{rpn}}$ is the ``current-squared'' spectral density of noise stemming from $\epsilon_\R{rpn}$, and (X=U,L), where we assume equal intensity sidebands  $I^{0}_\R{lo\otimes lsb}= I^{0}_\R{lo\otimes usb}$ for notational simplicity.

A useful quantity related to SNR is the ``minimum detectable cloud phase shift'' $\Delta\phi_\R{min,cl}$, defined implicitly from \eref{eq:SNR} by setting $\R{SNR(V_\R{SUSHI},}~\Delta\phi_\R{min,cl}')\equiv 1$ with $\C{S}_{I^{2}\R{rpn}}\equiv0$. It directly reveals how the two main sources of noise (namely,  the fundamental quantum uncertainty of the probe beam phase, and excess technical phase noise between sidebands) combine to limit the sensitivity of SUSHI:
\be
\label{eq:Deltaphimin}
\Delta\phi_\R{min,cl,SUSHI} = \sqrt{(\Delta\phi_\R{min,cl,SUSHI}')^{2} + \C{S}_{\phi^{2}\R{rpn}}}
\ee
for a measurement bandwidth of $B=1\R{Hz}$.  Here,  $\C{S}_{\phi^{2}\R{rpn}}$  is the two-sided phase noise power spectral density (units rad$^{2}$/Hz) of $\epsilon_\R{rpn}$, which may be directly measured via FFT analysis of the OPLL1 error signal.  We may relate $\C{S}_{\phi^{2}\R{rpn}}$ to $\C{S}_{I^{2}\R{rpn}}$ by a simple model.  We observe that SUSHI with phase-noisy sidebands is the same as SUSHI with phase-noiseless sidebands but with the probe beam passing through a fictitious noisy phase object (in series with the atom cloud).  In this picture the phase fluctuations of the fictitious object ``cause'' the observed excess technical phase noise between the sidebands having power spectral density  $\C{S}_{\phi^{2}\R{rpn}}$.  The effect of the fictitious noisy object on the SUSHI signal then parallels the effect of the real atomic cloud, inducing equal and opposite phase noise on each sideband, so that the single beam noise term should be identified as $\langle\epsilon_\R{fict}\rangle=\sqrt{\C{S}_{\phi^{2}\R{rpn}}}/2.$ The fictitious phase object model allows us to get the RF root-mean-squared noise current stemming from technical noise fluctuations by simply taking the (SUSHI signal) numerator of \eref{eq:SNR} and replacing $\Delta\phi_\R{cl}$ by $\langle\epsilon_\R{fict}\rangle$, arriving at
\be
\sqrt{\C{S}_\R{I^{2}\R{rpn}}} = \sqrt{\C{S}_{\phi^{2}\R{rpn}}}\kappa\sqrt{2\C{P}_\R{lo}}\sqrt{2\C{P}_\R{xsb}}
\ee
so that \eref{eq:SNR} may be computed completely from experimentally measured values.

\begin{figure}
\center
\includegraphics[width=3.3in]{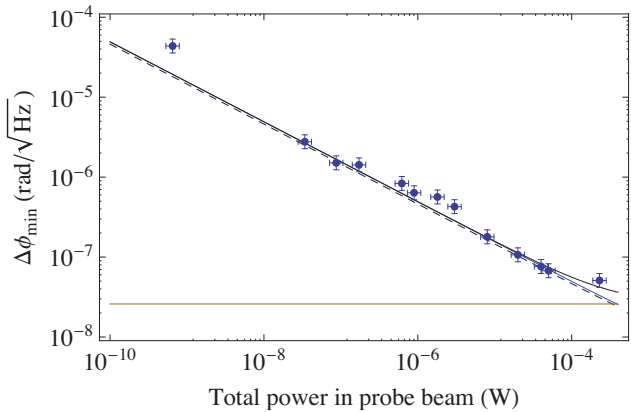}
\caption{SUSHI minimum detectable cloud phase shift versus total probe beam power.  The dashed black line is the fundamental limit for heterodyne phase detection of an atom cloud induced phase shift, using a noiseless detector of quantum efficiency $\kappa= 0.6$.  The blue line is a no-free-parameters calculation of the theoretical sensitivity of a SUSHI measurement using a probe beam composed of two equal power sidebands totaling the shown power, a 2.4 mW LO beam, and a detector having dark noise $\C{S}_\R{I^{2}det}=(17 \R{pA}/\sqrt{\R{Hz}})^{2}$.  The brown horizontal line at $37~\R{nrad}/\sqrt{\R{Hz}}$ is the sensitivity floor set by the  $\langle \epsilon_\R{rpn}\rangle=52~ \R{nrad}/\sqrt{\R{Hz}}$ residual phase noise between the USB and LSB fields. (\Sref{sec:detlimits} discusses why the former is smaller than the latter by the factor $1/\sqrt{2}$.)  The black curve is the uncorrelated combination of the two sources of phase noise.  Solid circles are our experimental determinations of the minimum detectable signal based on measurements of the phase noise floor of the final SUSHI signal, as described in Section \ref{sec:Experiment}.}
\label{fig:SUSHISNRdata}
\end{figure}

In \Fref{fig:SUSHISNRdata} we plot the theoretical minimum detectable phase shift for SUSHI with our particular LO beam power, detector noise, and residual phase noise, and compare to experimentally measured phase noise observed at various probe beam powers.  Details of the measurement are given in Section \ref{sec:Experiment}.  Here we comment that the data agree well with the theory, and reveal close to an SNR within 3dB of the quantum limit across more than 5 orders of magnitude variation in probe beam power.

\subsection{SUSHI at the standard quantum limit}
\label{sec:SUSHISQL}
In the bright local oscillator limit ($\C{P}_\R{lo}\gg \C{P}_{xsb},\C{S}_{I^{2}\R{det}}$), neglecting technical phase noise, and setting $\C{P}_\R{probe}=2\C{P}_\R{xsb}$, \eref{eq:SNR} becomes
\be
\label{eq:SNR(SUSHI,sql)}
\left[\R{SNR(V_{SUSHI})}\right]_{P_\R{lo}\rightarrow \infty}\sqrt{B}=\Delta \phi_\R{cl} \sqrt{2\kappa} \sqrt{\C{P}_\R{probe}}
\ee
Comparing this result to \eref{eq:SNRP}, we see that $K_\R{SUSHI,P_{LO}  \rightarrow\infty}=1$.  Curiously, the SUSHI measurement of an atom cloud---fundamentally a difference of \emph{two} independent heterodyne measurements of the cloud---has the same SQL SNR as \emph{one} heterodyne phase measurement of the same cloud using the \emph{same} probe power.  A moment's reflection reveals why this should be so.  The uncorrelated noises of the two measurements add in quadrature, increasing the noise by $\sqrt{2}$, but because the amplitude of the beat note photocurrent is proportional to the probe field \emph{amplitude}, not power, by dividing the probe power in half and making two measurements of $\Delta\phi_\R{cl}$ one can combine the measurements to make a total signal larger by the same factor $\sqrt{2}$.  It also follows that the minimum detectable phase shift for SUSHI in the bright LO limit is equal to the smallest allowed by quantum mechanics for for any heterodyne interference measurement of the phase shift of a classical probe laser:
\be\label{eq:SQLmindeltaphi}
\left[\Delta\phi_\R{cl,min,SUSHI}\right]_{P_\R{lo}\rightarrow \infty}=\sqrt{\frac{B}{2\kappa\C{P}_\R{probe}}}.
\ee

Finally, we remark that while the condition of opposite phase shifts of the two sidebands is not required for operation of the SUSHI scheme per se, it is essential if SUSHI is to detect atom cloud induced phase shifts at the standard quantum limit for a fixed total probe beam power.

\section{SUSHI for Minimally Destructive Measurements}
Here we calculate the relationship between SNR and spontaneous scattering for SUSHI detection in the $P_\R{lo}\rightarrow \infty$ limit, demonstrating that $\R{SNR}\propto\sqrt{\Gamma_{sc}}$ and revealing an important dependence on cloud shape.  We treat first a gas of two-level atoms (one ground state and one optically excited state); we then generalize these results to multilevel alkaline atoms.

\subsection{SUSHI with 2-level atomic gases}
\label{sec:SUSHIwith2levelgasses}
Consider a right cylinder of cross sectional area $A$ and length $l$ filled with a homogenous gas of $N$ 2-level atoms (i.e., uniform density $n=N/l A$, uniform column density $n_\R{col}=nl=N/A$). Model the SUSHI probe beam as a plane wave of intensity $I=2 P_\R{xsb}/A$, where $P_\R{xsb}$ is the power in one of the two (assumed equal intensity) sideband fields.  From \eref{eq:2lphaseshift}, with $\delta_\R{xsb}\equiv(\omega_\R{xsb}-\omega_{0})/\Gamma\gg 1$,
\begin{align}
\label{eq:SUSHI2L}
\Delta\phi_\R{SUSHI}^{(2l)}&=\Delta\phi_\R{cl,lsb} - \Delta\phi_\R{cl,usb} \nonumber\\
&=\frac{N \sigma_{0}}{4 A}\left(\frac{1}{\delta_\R{lsb}} + \frac{1}{\delta_\R{usb}}\right) = \frac{N \sigma_{0}}{2 A \delta}
\end{align}
where the last equality follows if the sidebands are symmetrically detuned about the atomic transition ($\delta_\R{usb}=-\delta_\R{lsb}=\delta$).  For any achievable sideband splitting, and for equal sideband powers, the ratio of the SUSHI phase shift \eqref{eq:SUSHI2L} to spontaneous scattering rate \eqref{eq:SUSHISLGamma} is maximized for symmetric detuning of the sidebands about the resonance.  By a simple extension of the ideas of \Sref{sec:SUSHISQL} it can easily be shown that, for a fixed total probe intensity, and for symmetric sideband detuning, the SQL SNR for SUSHI is maximized for equal power sidebands. Therefore, it is both reasonable and notationally convenient to make these assumptions here.

\begin{figure}
\center
\includegraphics*[width=3.2in]{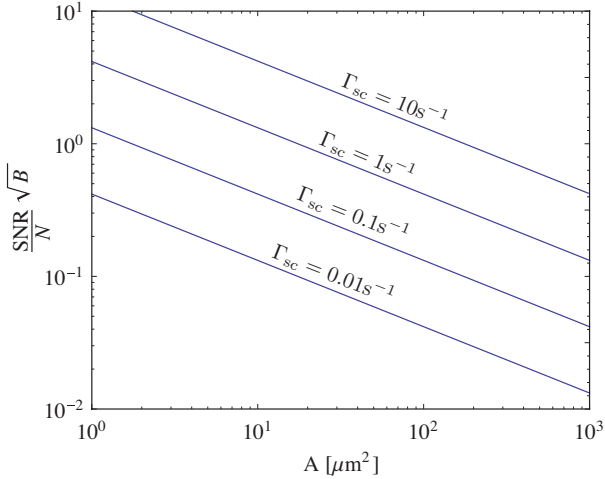}
\caption{\label{fig:SQLSNR}Plot of \Eref{eq:SQL2levelSNR}, the (per atom) signal-to-noise root bandwidth product versus cross-sectional area of the cloud for a probe on the 780 nm D2 line of $^{87}$Rb, in the 2-level limit.  The curves correspond to four different choices of (per atom) spontaneous scattering rate.  A quantum efficiency of $\kappa=0.6$ was assumed for the calculations.}
\end{figure}

The probe light causes spontaneous scattering in the atomic cloud at a (per atom) rate of
\be
\label{eq:SUSHISLGamma}
\Gamma_{sc}^{(2l)}=\frac{1}{(t_{3})^2}\times\frac{\sigma_{0}}{4 A}\left(\frac{\C{P}_\R{lsb}}{\delta^{2}_\R{lsb}} + \frac{\C{P}_\R{usb}}{\delta^{2}_\R{usb}}\right) = \frac{\C{P}_\R{probe} \sigma_{0}}{(t_{3})^2 4 A \delta^{2}},
\ee
where $(t_{3})^2$ is the transmission coefficient of beamsplitter BS3 (see \Fref{fig:opticalcircuit}), and $\C{P}_\R{probe}$ is the total probe power reaching detector D2.  We can use this expression to re-write \eref{eq:SNR(SUSHI,sql)} as an explicit function of the scattering rate:
\be
\label{eq:SQL2levelSNR}
\frac{\left[\R{SNR(V_{SUSHI})}\right]^{(2l)}_{P_\R{lo}\rightarrow \infty}}{N}\sqrt{B}=\sqrt{\frac{2\kappa\sigma_{0}\Gamma_{sc}^{(2l)}}{A}}
\ee
for $t_{3}\rightarrow1$.   We find that the $\R{SNR}\sqrt{B}$ product (per atom) for measuring a 2-level atomic gas at a given (per atom) spontaneous scattering rate is proportional to $\sqrt{\Gamma_{sc}}$, and that the proportionality constant is a dimensionless number that, apart from the detector quantum efficiency, depends only on the ratio of the resonant light-scattering cross section to the cross-sectional area of the cloud.  In other words, a SUSHI measurement performed with any combination of probe power and detuning that generates the same spontaneous scattering has the same $\R{SNR}\sqrt{B}$.  However, one can achieve a \emph{higher} standard quantum limited SNR$\sqrt{B}$ for same (per atom) scattering rate for prolate clouds as compared to round or oblate clouds, for the same number of atoms.  (This dependence is not unique to SUSHI, but is generally true of dispersive probes of atomic clouds.) In \Fref{fig:SQLSNR} we plot several contours of $\R{SNR}\sqrt{B}$ (per atom) versus cross-sectional area for different (per atom) scattering rates, for a probe laser of $\lambda=780$~nm and $\sigma_{0}=3\lambda^2/2 \pi$.

\subsection{SUSHI with multi-level atomic gases}
\label{sec:m}
We now generalize our treatment of SUSHI to multilevel atoms; specifically, to alkaline atoms having two hyperfine ground states separated by a microwave frequency interval, and two excited state manifolds (D1 and D2) separated by an optical frequency interval.  We present the results of 2nd order time-independent perturbation theory calculations of spontaneous scattering rates and phase shift signals for SUSHI detection of atoms initially prepared in one ground state hyperfine level, but otherwise unpolarized  (i.e., equal $m_{F}$ populations), for probe laser wavelengths detuned (by microwave frequencies) from either the D1 or D2 line.  Since the results depend on many experimental parameters, for clarity we consider the scaled (per atom) SUSHI phase shift signal:
$$\widetilde{\Delta\phi}_{\R{SUSHI}} \equiv \frac{A}{N} \Delta\phi_{\R{SUSHI}},$$
where $A$ is the cross-sectional area of a cylindrical cloud and $N$ is the total atoms in the cloud;
and the scaled (per atom) scattering rate, 
$$\widetilde{\Gamma}_{\R{sc}} \equiv \frac{A}{\C{P}_\R{probe}}\Gamma_{sc} =\frac{A}{P_\R{probe}/\hbar \omega}\Gamma_{sc},$$
where $P_\R{probe}$ is the total power \emph{incident on the atoms}.  Both scaled quantities have dimensions of [length$^{2}$].

The multilevel generalization of \eref{eq:SUSHI2L} is
\begin{align}
\label{eq:DeltaphiminML}
\widetilde{\Delta\phi}_{\R{SUSHI}}(Dn,F) = \sigma_{0}\sum_{F'} T.S.(Dn, F, F')  \times \nonumber\\
\left(\frac{\delta_{\R{lsb}}(Dn,F,F')}{1 + 4~[\delta_{\R{lsb}}(Dn,F,F')]^{2}}- \frac{\delta_{\R{usb}}(Dn,F,F')}{1 + 4~[\delta_{\R{usb}}(Dn,F,F')]^{2}}\right),
\end{align}
where $F$ is a ground hyperfine state, $F'$ is an optically excited hyperfine state,  $\delta_\R{xsb}(Dn,F,F') \equiv \frac{\nu(\R{xsb})-\nu(F\rightarrow F')}{\Gamma(Dn)}$ is the detuning of the XSB ($X=U,L)$ sideband from the $F\rightarrow F'$ transition in units of the natural linewidth of the $Dn$ ($n=1,2)$ line, and $T.S.(Dn, F, F')$ is the appropriate (angular) averaged transition strength~\cite{Metcalf:1999}.
The multilevel generalization of \eref{eq:SUSHISLGamma} is
\begin{align}
\label{eq:Gammam}
\widetilde{\Gamma}_{\R{sc}}(Dn,F) = \frac{1}{(t_{3})^{2}}\times\sigma_{0}\sum_{F'} T.S.(Dn, F, F')  \times \nonumber\\ \left(\frac{1}{1 + 4~[\delta_{\R{lsb}}(Dn,F,F')]^{2}}+\frac{1}{1 + 4~[\delta_{\R{usb}}(Dn,F,F')]^{2}}\right).
\end{align}

\begin{figure*}
\center
\includegraphics*[width=7.2 in]{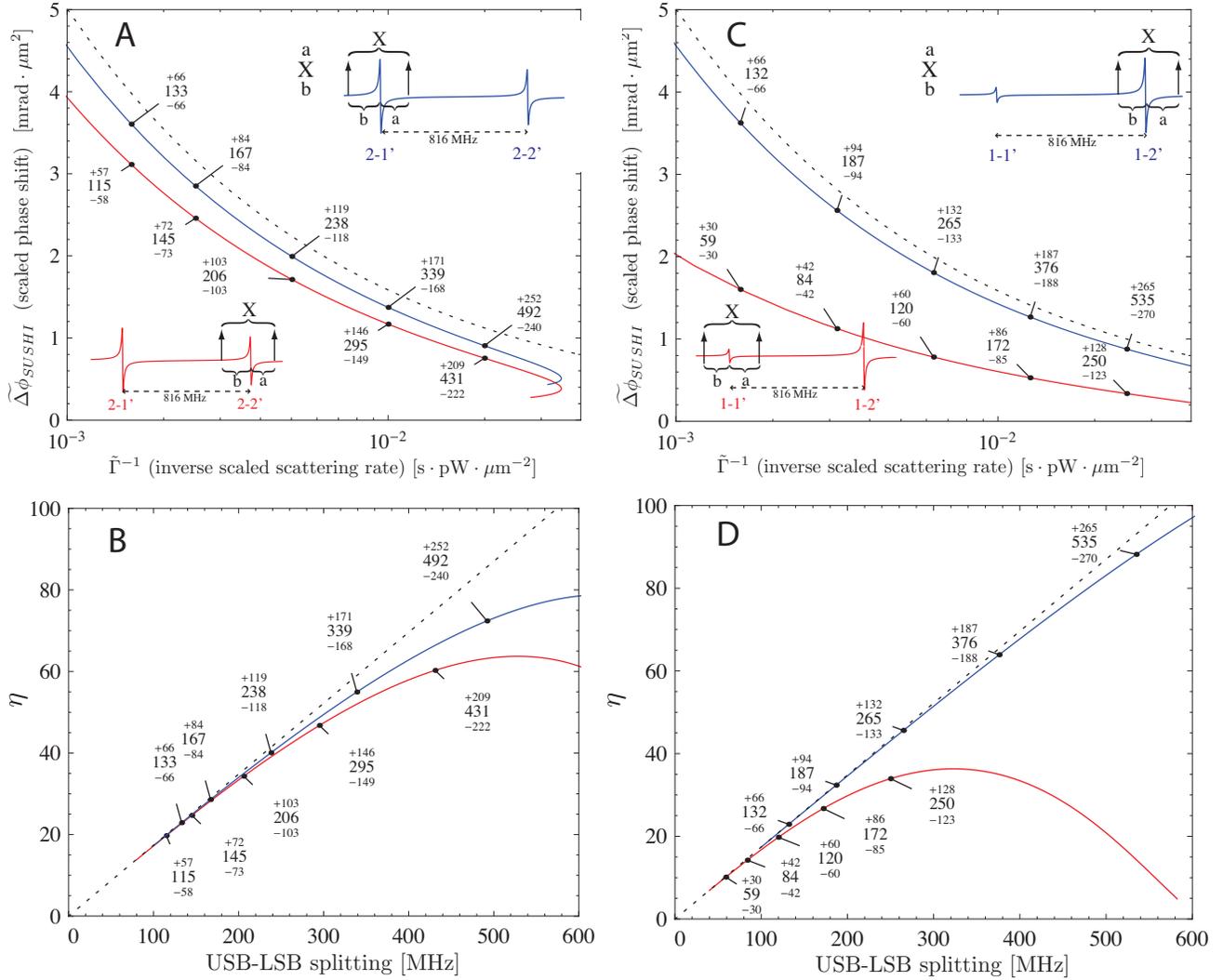}
\caption{\label{fig:D1} Calculated SUSHI performance metrics for the \emph{internal bracketing} detuning configuration on the  $^{87}$Rb D1 line.  {\bf[4A, main]:}  A plot of scaled phase shift versus inverse scattering rate.  Blue curve is for sidebands bracketing the $(F=2 \rightarrow F'=1)$ transition, as depicted in the upper right inset; red curve is for sidebands bracketing the $(F=2 \rightarrow F'=2)$ transition, as depicted in the lower left inset.  Black, dashed curve is a calculation valid for resolved fine structure but completely unresolved hyperfine structure, provided for reference.     {\bf [4A, insets]} LSB is the leftmost arrow, USB the rightmost.  $X$ is the total frequency interval between sidebands; $a$ and $b$ are the detuning of the LSB and USB, respectively, from the nearest allowed transition.  {\bf[4B:]} A plot of efficiency versus sideband splitting.  Colors and points match those of 4A;  black, dashed curve is $\eta_{2l}$, the two level atom result.   {\bf [4C,D]:} Same as [4A,B], but for atoms in the $F=1$ hyperfine ground state. Blue curves are for sidebands bracketing the $F=1 \rightarrow F'=2$ transition; red curve is for sidebands bracketing the $F=1 \rightarrow F'=1$ transition.}
\end{figure*}

\begin{figure*}
\center
\includegraphics*[width=3.7 in]{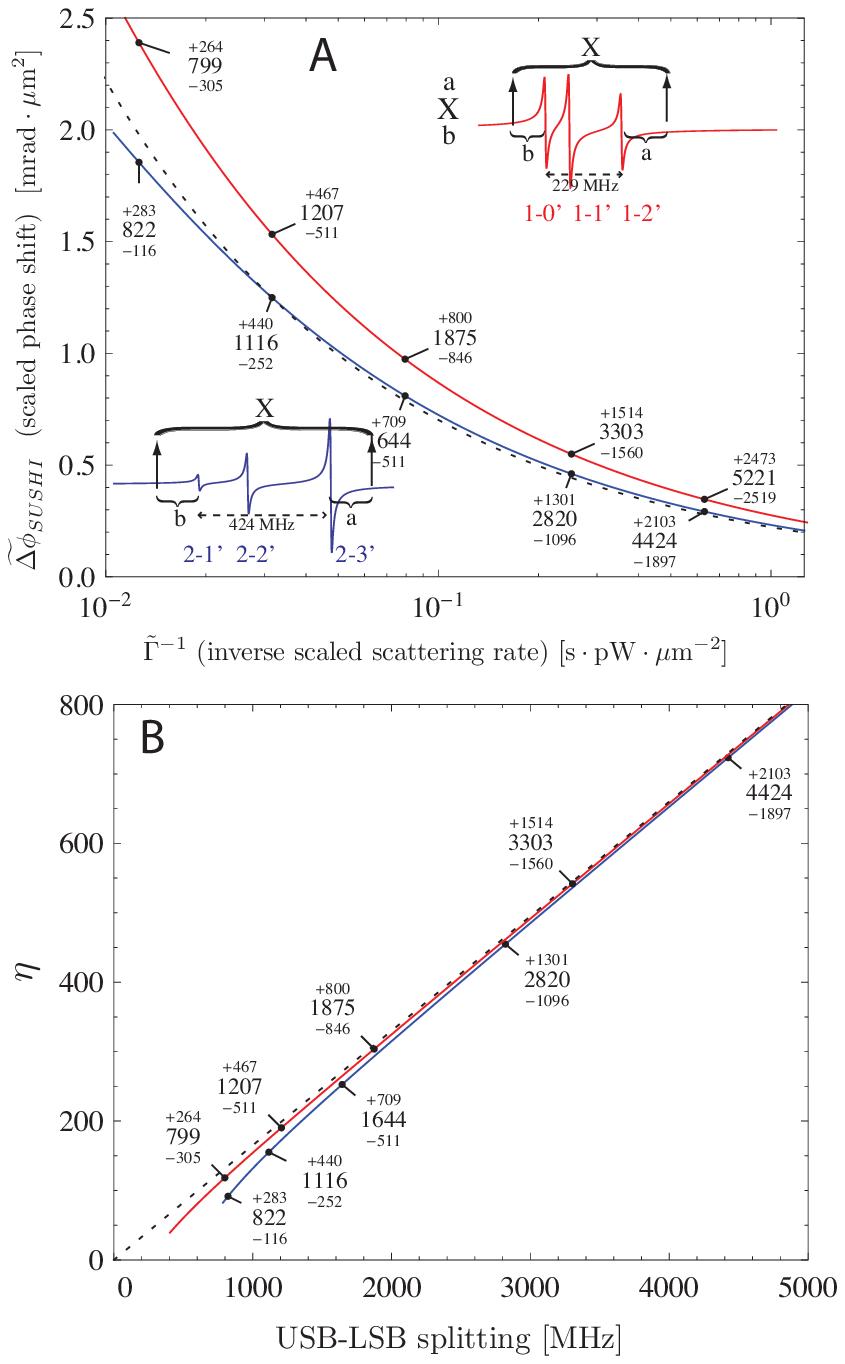}
\caption{\label{fig:D2} Similar to Fig. 4, but for the \emph{external bracketing} configuration on the  $^{87}$Rb D2 line, as described in the text. {\bf[5A]} Red curve is for unpolarized atoms are in the $F=1$ hyperfine ground state, LSB below the $F=1 \rightarrow F'=0$ transition and USB above the $F=1 \rightarrow F=2'$ transition, as depicted in the upper right inset.  Blue curve is for unpolarized atoms in the $F=2$ hyperfine ground state, LSB below the $F=2 \rightarrow F'=1$ transition and USB above the  $F=2 \rightarrow F'=3$ transition, as depicted in the lower left inset.  Black, dashed curve is a calculation valid for resolved fine structure but completely unresolved hyperfine structure, provided for reference. {\bf [5B]} The red (blue) curve is for atoms in the $F=1 (F=2)$ ground state, respectively.}
\end{figure*}

A useful performance metric is the ratio  $\eta \equiv \widetilde{\Delta\phi}_\R{SUSHI}/\widetilde{\Gamma}_{sc}$, which can be understood as the (per atom) differential phase shift between the LSB and USB sidebands for each scattered probe photon.  For SUSHI on a 2-level atom with equal power sidebands split symmetrically from the optical resonance by $\pm\delta$, $\eta^{(2l)}=2\delta$.  For alkaline atoms $\eta$ approaches this value when $\delta$ is much less or much greater than the width of the nearest excited state manifold, but can deviate significantly at intermediate detunings (see \Fref{fig:D1} and \Fref{fig:D2}).

We apply this formalism to the specific case of $^{87}$Rb.  The $5S_{1/2}$ ground state consists of two hyperfine levels $(F=1,2)$ split by 6.8 GHz. Each level is connected by a 795 nm D1 line to the $5P_{1/2}$ manifold consisting of two hyperfine levels separated by 816 MHz; and by a 780nm D2 line to the  $5P_{3/2}$ manifold consisting of four hyperfine levels with an average separation of 165 MHz.  We consider two possible sideband detuning configurations: an \emph{internal bracketing} configuration on the D1 line, suited to small detunings,  in which one of the sidebands is located between two excited state levels; and an \emph{external bracketing} configuration on the D2 line, suited to large detunings, in which both sidebands are located completely outside of the exited state manifold.  In both cases we consider only detunings larger than the natural linewidth but smaller than 6.8 GHz ground state hyperfine splitting.

In \Fref{fig:D1} we plot $\widetilde{\Delta\phi}_{\R{SUSHI}}$ versus $\widetilde{\Gamma}_{sc}^{-1}$, and $\eta$ versus the total sideband splitting $X=|\nu_{usb}-\nu_{lsb}|$, for the \emph{internal bracketing} configuration implemented on the D1 line.  For these calculations, atoms are assumed to be in the $F=2$ hyperfine ground state, equally distributed among magnetic sublevels.  A few selected operating points are labeled with the values of sideband frequencies corresponding to equal spontaneous scattering rates.  The 816 MHz hyperfine splitting of the $5P_{1/2}$ manifold is compatible with sideband splittings  less than $500$~MHz.  For larger splittings $\eta$ begin to turn down when the interior sideband approaches the other excited state (see Figures \ref{fig:D1}B and 4D).  In \Fref{fig:D2} we present similar calculations for the \emph{external bracketing} configuration, implemented on the D2 line.  Here the sidebands bracket all allowed transitions to the $5P_{3/2}$ manifold.  As can be seen in the insets to \Fref{fig:D2}A, the sideband splitting $X$ is greater than the sum of the detuning of each sideband to its nearest allowed transition (i.e., $X>a+b$).  The ultimate performance of the external bracketing configuration is, of course, superior to internal bracketing, but with the burden of higher frequency heterodyne detection circuitry. 

\section{Experiment}
\label{sec:Experiment}

\begin{figure}
\center
\includegraphics*[width=3.3 in]{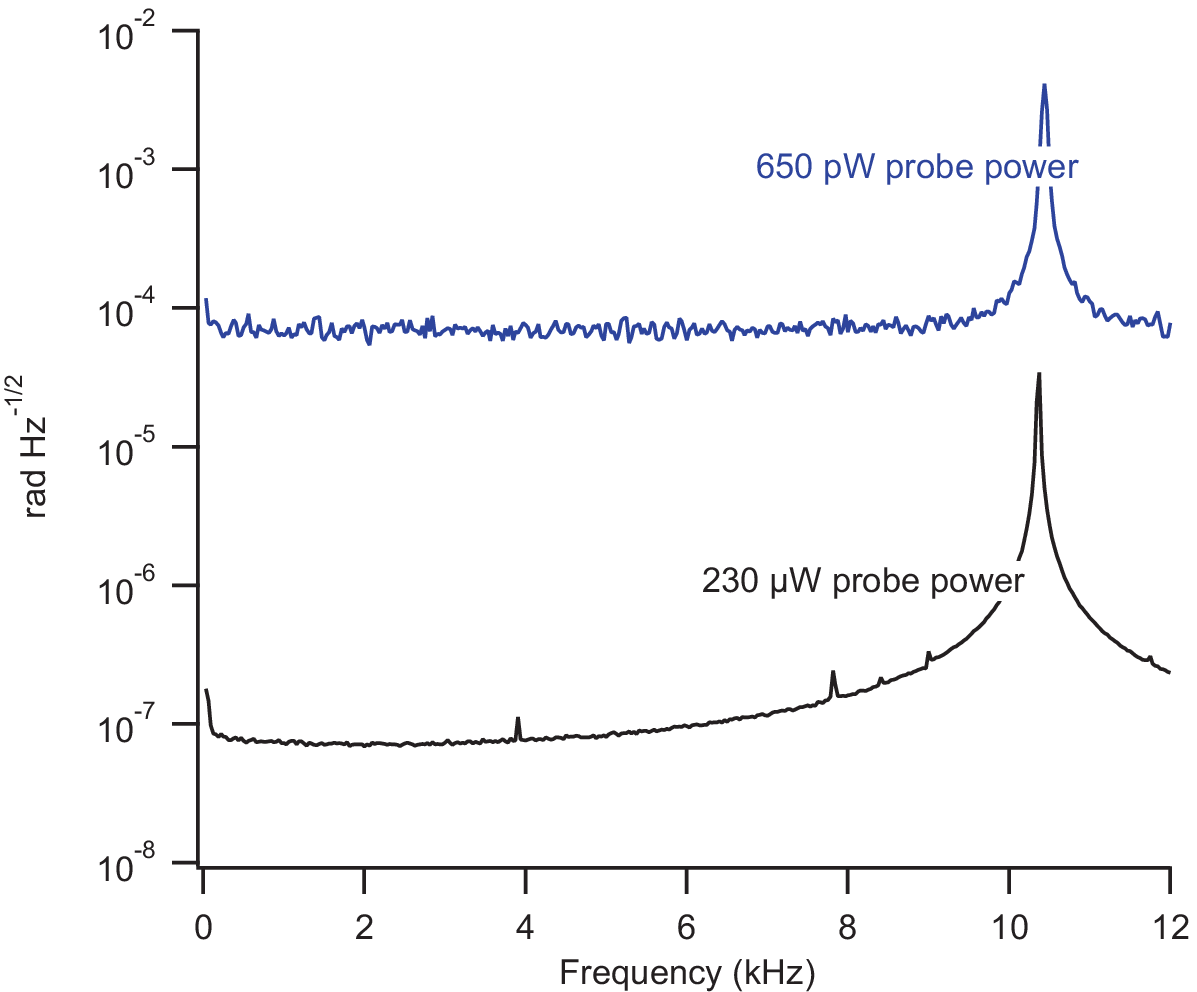}
\caption{\label{fig:SUSHIpeak}Example phase noise spectra for SUSHI detection.  Gain peaks in the OPLLs are observed around 10 kHz. Upper, blue (lower, black) trace is the observed noise floor for SUSHI detection with a 650 pW (230 $\mu$W) probe beam, respectively.  The noise values used to infer the minimum detectable atomic phase signal of \Fref{fig:SUSHISNRdata} are taken at 1.5 kHz from data sets like these, measured at the various probe powers.}
\end{figure}

Here we present measurements of the SUSHI minimum detectable phase signal versus probe power, and describe additional details of our apparatus.  In our implementation of SUSHI we derive all needed optical fields from a single 220 mW, fiber coupled, diode laser source.  The source laser is shot noise limited at the RF frequencies used for heterodyne phase detection.  While we found that our PFD phase detectors strongly reject baseband amplitude fluctuations, nevertheless our results improved (i.e, our noise floor was lowered) when we added a single-stage amplitude stabilization circuit to the source laser, which clamped intensity fluctuations from 0 to 10 kHz to within 5~dB of the shot noise limit.  This pre-stabilization stage was in place for all of the results presented here.  The USB and LSB fields each are made starting from 7 mW of source light, frequency shifted by $+$160 MHz and $-$220 MHz, respectively, in double-passed AOMs.  As much as 230 $\mu$W of probe light can be directed onto the atoms.  The LO field derives from 150 mW of source light, frequency shifted by $-$75 MHz in a single-pass AOM.  75 mW of LO power illuminates beamsplitter BS3, yielding 2.4 mW incident on detector D2.

The AOMs used to steer optical phases in our two OPLLs are driven by low phase-noise VCOs, battery powered, and shielded from EMI by a metal enclosure.  Extreme care was taken to avoid ground loops in the RF synthesis and amplification chain.  The analog PID circuits steering the VCO frequencies are built from ultra-low-noise op-amps (Analog Devices ADA4075-2), also battery powered and located in the same enclosure as the VCO's.  The fixed 80 MHz source for the LSB AOM, and the 39 MHz, 145 MHz and 117.5 MHz phase reference signals for the the LSB$\otimes$USB, LO$\otimes$LSB, and LO$\otimes$USB phase detection circuits are coherently synthesized by DDS ICs (Analog Devices AD9599) directly clocked by a low phase noise master synthesizer (Stanford Research Systems SG384).

We measure the residual phase noise $\epsilon_\R{rpn}$ of OPLL1 (between the USB and LSB fields) with a separate, out-of-loop PFD and a low frequency FFT spectrum analyzer.  The observed closed-loop residual phase noise of 52 nrad/$\sqrt{\R{Hz}}$ from 700~Hz to 8~kHz is limited approximately equally by photodetector RF dark noise, photon RF shot noise, and baseband excess amplitude noise of the probe laser.  The identical high-speed optical detectors D1 and D2 have a measured NEP=$17~\R{pW/\sqrt{Hz}}$---equivalent to the shot noise of a 5mW beam---at the 145 MHz, 235 MHz, and 380 MHz frequencies at which we make heterodyne phase measurements.  The PFD circuits for OPLL1 (Analog Devices AD4113) and OPLL2 (Analog Devices AD4002) are insensitive to technical amplitude noise on the input signals, so that our measured probe beam intensity noise of 5~dB above shot noise generates only  37 nrad/$\sqrt{\R{Hz}}$ of excess phase noise; the remainder of the 52 nrad/$\sqrt{\R{Hz}}$ total residual phase noise comes from intrinsic photon shot noise and detector dark noise.

In \Fref{fig:SUSHISNRdata} we show experimentally determined minimum detectable atom cloud induced phase shifts for various probe powers between 650~pW and 230~$\mu$W.  These are calculated from FFT measurements of the phase noise floor of the final SUSHI signal (i.e., the LO$\otimes$USB phase detector).  For all measurements the amplitude of the LO$\otimes$USB RF beat note was kept at a constant -10~dBm as the probe power was changed, by either amplification or attenuation in the RF domain.  We found the SUSHI noise floor to be flat from 60~Hz to 8~kHz.  The data show detection sensitivity within $\sim$1 dB of the quantum limit for our q.e.$=0.6$ photodetector for probe powers from 86~nW to  35 $\mu$W, and within $\sim$3 dB from 650 pW to 230 $\mu$W, and thus compare extremely well with the no-free parameters calculation of \eref{eq:Deltaphimin} (blue line of \Fref{fig:SUSHISNRdata}).   Our technical phase noise floor of 37 nrad/$\sqrt{\R{Hz}}$ (brown line of \Fref{fig:SUSHISNRdata}) is sufficiently low that only measurements made with our (brightest achievable) 230 $\mu$W probe beam begin to show signs of deviation from the fundamental quantum limit.  Vertical error bars in \Fref{fig:SUSHISNRdata} are dominated by systematic uncertainties in calibration of the FFT spectrum analyzer; horizontal error bars reflect a systematic uncertainty in the calibration of the optical power meter used to measure the probe beam power

\section{Conclusions}
In this work we have introduced SUSHI, a new technique for dispersive detection of atomic clouds capable of operating in the deeply quantum noise limited regime using off-the-shelf detectors, closely approaching the maximum sensitivity allowed for any heterodyne phase measurement using classical light.  Our novel active subtraction scheme strongly rejects phase noise due to mechanical vibrations, establishing a technical noise floor for the SUSHI measurement of 37 nrad/$\sqrt{\R{Hz}}$ without adjustable mirrors or auxiliary far-detuned lasers, and even for large, time-dependent phase signals.  In this work we have demonstrated near quantum limited sensitivity for probe beam powers as low as 650 pW.  Operation at even lower probe powers should be possible if the PFDs were replaced by traditional analog mixers, as we have discovered that digital PFDs malfunction when presented with RF signals having SNR$<$10.

We remark that SUSHI could be used to implement the protocol of Appel et al.~\cite{Appel:2009oq} for producing spin squeezed states via single-pass QND measurements.  To do this, the two probe components must be separated in frequency by $8.4$GHz (for $^{87}$Rb).   In our current apparatus the maximum achievable LSB-USB splitting is $\sim360$~MHz.  However, a straightforward modification of the probe beam synthesis chain should be capable of producing LSB-USB splittings of  $\sim$10~GHz.  Specifically, the USB and LSB RF frequency AOMs would be replaced by microwave frequency (but low efficiency) versions, and the dim shifted light coherently amplified by injection seeding separate diode lasers as in Ref.~\cite{Durfee:2006li}.  We note that PFDs and photoreceivers operating at 10GHz, and that otherwise meet the essential performance criteria for SUSHI, are commercially available.

In the future we plan to use SUSHI to make make real-time observations of Rabi flopping of atoms for fundamental studies of measurement induced decoherence in mesoscopic systems, and to implement a novel protocol for in-situ, non-destructive magnetometry of Bose-Einstein condensates.

This material is based upon work supported by the U. S. Army Research Office under grant number W911NF-09-1-0179.

\appendix
\section{Quantum limited dispersive detection with noisy detectors; 1- and 2-arm heterodyne techniques compared.}
\label{sec:1vs2}
We compare the performance of 1-arm to  2-arm heterodyne interferometry for making weak dispersive measurements. 
We calculate the SNR for sensing an atom-cloud induced relative optical phase shift $\Delta \phi$ between two optical fields of powers $P_{1}$ and $P_{2}$ and frequencies $\omega_{1}\neq\omega_{2}$.  In 1-arm detection both fields pass through the cloud; in 2-arm detection $P_{2}$ does not. In both cases the fields are overlapped on a fast photodetector, and the phase of the resulting RF-frequency photocurrent is compared to an ideal RF reference in an electronic phase detector. In both cases the SNR of the heterodyne photocurrent, and thus of the (baseband) signal from the phase detector, is given by
\be
\label{eq:SNR1and2}
\R{SNR}\sqrt{B} =  \Delta\phi\frac{\sqrt{2\kappa \C{P}_1}\sqrt{2\kappa \C{P}_2}}{\sqrt{2\kappa \C{P}_1 + 2\kappa \C{P}_2 + \C{S}_{I^{2}det}}}
\ee
for a photodetector of quantum efficiency $\kappa$ and dark noise ``current-squared'' spectral density $\C{S}_{I^{2}det}$  (units of (electrons/s)$^{2}$/Hz).  The first two terms under the radical of the denominator are photocurrent shot noise ``current-squared'' spectral densities, the pre-factors of 2 coming from the Schottky shot noise formula when the characteristic detector averaging time $\tau$ is expressed in terms of an effective measurement bandwidth $B\equiv1/2\tau$.   We do not here consider technical noise due to mechanical disturbances of beam paths or to excess laser intensity noise.

For 1-arm measurement schemes the two fields are combined into a single probe beam of power $P_\R{probe}=P_1+P_2.$  For convenience we write the powers in terms of a balance factor $f$: $P_{1}=f P_\R{probe}, P_{2}=(1-f) P_\R{probe}$, for $0<f<1$.  For 2-arm measurement schemes we identify the first field with the probe beam, $P_1=P_\R{probe}$ and the other field with the ``local oscillator'' (LO) beam which bypasses the atoms: $P_2 = P_\R{lo}$.  With minor rearrangement \eref{eq:SNR1and2} takes the form of \eref{eq:SNRP}:
\be
\R{SNR}\sqrt{B} = (K 'K'') \Delta \phi \sqrt{2\kappa}\sqrt{\C{P}_\R{probe}},
\ee
The terms $K'$ and $K''$  have different dependencies on experimental parameters for 1-arm schemes as compared to 2-arm schemes.  Specifically, $K'$ is given by 
\begin{align}
K'_\R{1 arm} =& \sqrt{f(1-f)}\\
K'_\R{2 arm} =& \left(1+\frac{\C{P}_\R{probe}}{\C{P}_\R{lo}}\right)^{-1/2}
\end{align}
$K'_\R{1 arm}$ takes its maximum value of 1/2 when the probe power is balanced between the two fields $(f=1/2)$, whereas $K'_\R{2 arm}\rightarrow 1$ in the bright LO limit $(\C{P}_\R{lo}\gg\C{P}_\R{probe})$.

The main advantage to 2-arm methods is revealed in the $K''$ factor, given by 
\begin{align}
K''_\R{1 arm} =& \left(1+\frac{\C{S}_{I^{2}\R{det}}}{2\C{P}_\R{probe}}\right)^{-1/2}\\
K''_\R{2 arm} =& \left(1+\frac{\C{S}_{I^{2}\R{det}}}{2\C{P}_\R{lo}+2\C{P}_\R{probe}}\right)^{-1/2}
\end{align}
$K''_\R{1 arm}\rightarrow 1$ only for  $\C{P}_\R{probe}\propto\sqrt{\delta^{2}\Gamma_{sc}}\gg \C{S}_{I^{2}\R{det}}$; i.e., by operating with a  large probe beam detuning $\delta$ and/or large spontaneous scattering rate $\Gamma_{sc}$, as required to overcome detector dark noise.   This means that 1-arm heterodyne measurements that are simultaneously weak (small $\Gamma_{sc}$) and resolve closely spaced levels (small $\delta$) cannot be quantum noise limited with any currently available detector.   On the other hand $K''_\R{2 arm}\rightarrow 1$ even for $\delta\rightarrow 0$ and $\Gamma_{sc} \rightarrow 0$ due to the presence of  $\C{P}_\R{lo}$ in the denominator, which can be increased with no effect on  the spontaneous scattering rate.

\begin{figure}
\center
\includegraphics[width=3.2 in]{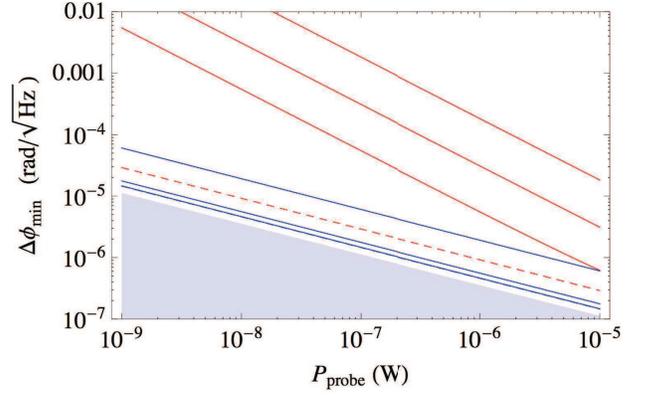}
\caption{Minimum detectable phase shift for a single heterodyne dispersive phase measurement, versus probe beam power, for 1- and 2-arm configurations.  Red lines are 1-arm detection with $f=1/2$ (see text); blue lines are 2-arm detection with $P_\R{lo}=2.4$~mW.  Each color group shows curves for three $\kappa=.6$ photodetectors having dark noise levels of $\R{N.E.P.}=3, 17, \mathrm{~and~} 100$~pW/$\sqrt{\R{Hz}}$ (lowest to highest curves on the graph).  The noiseless detector limit for each type of measurement is shown in dashed lines of the same color.  The top of the shaded region is the quantum limit for unit quantum efficiency, noiseless detection of a classical probe.  A probe laser wavelength of 780 nm is assumed.}
\label{fig:SNR1vs2}
\end{figure}

In \Fref{fig:SNR1vs2} the minimum detectable phase shift  ($\R{SNR}(\Delta \phi_{min})\equiv1)$ is plotted versus probe power for 1-arm and 2-arm detection schemes, for a selection of representative off-the-shelf photodetector noise metrics.

\section{Phase shift and attenuation of a weak, off resonant laser crossing a gas of 2-level atoms.}
\label{app:2level}
Here we summarize, for convenience and to establish notation, the well known results for the phase shift and attenuation of a weak, off resonant laser beam passing through a gas of 2-level atoms.  We draw from the treatments of Refs.~\cite{Ketterle:1999hc,Grimm:2000tp,Born:1999ws}.

A  scalar electromagnetic field of frequency $\omega$ crossing a homogenous, dispersive, and weakly absorptive medium of complex polarizability $\alpha$  and thickness  $\ell$ acquires a complex phase of the form
\be\label{eq:AppendixA1}E(x=\ell,t)=E_0\exp{[ - i \omega t + i\R{Re}(\kappa )\ell-\R{Im}(\kappa )\ell]}\ee
for the complex wave number 
\be\label{eq:AppendixA2}\kappa=\frac{\omega}{c}  \sqrt{1+\frac{n \alpha}{\epsilon_0}}\approx \frac{\omega}{c}  \left(1+\frac{n\alpha}{2\epsilon_0}\right),\ee
where the last approximation is valid for $n\alpha \ll 1$, as is well satisfied for a dilute atomic gas.

For a gas of 2-level atoms (transition frequency $\omega_{0}$, natural linewidth $\Gamma$) weakly driven (i.e., neglecting saturation effects) by a laser of frequency $\omega$, the complex polarizability is
\be\label{eq:AppendixA3}\alpha=6 \pi \epsilon_{0} c^3 \frac{\Gamma/\omega_{0}^{3}}  {\omega_{0}^{2} - \omega^{2}- i (\omega^{3}/\omega_{0}^{2})\Gamma}.\ee
In the rotating wave approximation, and writing the laser detuning in units of the natural linewidth $\delta\equiv(\omega-\omega_{0})/\Gamma,$ we have
\be\label{eq:AppendixA4}\R{Re}(\alpha ) = -\frac{12 \pi \epsilon_{0} c^3}{\omega^3} \frac{\delta}{1+ 4 \delta ^2},\ee
and from the Kramers-Kronig relations
\be\label{eq:AppendixA5}\R{Im}(\alpha )=-\frac{1}{2 \delta}\R{Re}(\alpha ).\ee

Using \eref{eq:AppendixA2}-\eref{eq:AppendixA5} we can express the complex phase in equation~\eref{eq:AppendixA1} in terms of the on-resonance light scattering cross section  $\sigma_0 \equiv \frac{3 \lambda ^2}{2 \pi }$ and the column density $n_\R{{col}}=n\ell$: 
\be \label{eq:AppendixA6}E(x=\ell,t)=E_0 \exp{\left[-i \omega t + i\omega (\ell/c)-i \Delta\phi -A \ell\right]}\ee
where 
\be\label{eq:2lphaseshift}\Delta\phi \equiv n_{col} \sigma_0 \frac{\delta}{1+ 4 \delta ^2}\ee  is the phase shift induced in the field by the gas, and
\be A\equiv\frac{1}{2}n_{col} \sigma_0 \frac{1}{1+ 4 \delta ^2}\ee is the amplitude attenuation exponent for the field due to scattering by the gas.

For a focussed gaussian laser beam crossing a spherical gaussian cloud of peak density $n_{0}$, in the case where the $1/e^{2}$ intensity radius $r_{0}$ of the beam is equal the same measure of the cloud density distribution, and for physical quantities averaged over the beam profile, one can make the simple replacement $n_{col} \rightarrow n_{0} \sqrt{\pi} r_{0}/2$ in the above formulae.


\begin{thebibliography}{10}
\newcommand{\enquote}[1]{``#1''}

\bibitem{Hope:2005kx}
J.~Hope and J.~Close, \enquote{General limit to nondestructive optical
  detection of atoms,} Phys. Rev. A \textbf{71}, 043822 (2005).

\bibitem{Andrews:1996dq}
M.~Andrews, M.~Mewes, N.~van Druten, D.~Durfee, D.~Kurn, and W.~Ketterle,
  \enquote{Direct, nondestructive observation of a bose condensate,} Science
  \textbf{273}, 84--87 (1996).

\bibitem{Ketterle:1999hc}
W.~{Ketterle}, D.~S. {Durfee}, and D.~M. {Stamper-Kurn}, \enquote{{Making,
  probing and understanding Bose-Einstein condensates},} eprint
  arXiv:cond-mat/9904034  (1999).

\bibitem{Liu:2009vo}
Y.~Liu, E.~Gomez, S.~Maxwell, L.~Turner, and E.~Tiesinga, \enquote{{Number
  Fluctuations and Energy Dissipation in Sodium Spinor Condensates},} Phys.
  Rev. Lett. \textbf{102}, 225301 (2009).

\bibitem{Windpassinger:2008cr}
P.~J. Windpassinger, D.~Oblak, U.~B. Hoff, J.~Appel, N.~Kj{\ae}rgaard, and
  E.~S. Polzik, \enquote{Inhomogeneous light shift effects on atomic quantum
  state evolution in non-destructive measurements,} New Journal of Physics
  \textbf{10}, 053032 (2008).

\bibitem{Windpassinger:2008tg}
P.~J. Windpassinger, D.~Oblak, P.~G. Petrov, M.~Kubasik, M.~Saffman, C.~L.~G.
  Alzar, J.~Appel, J.~H. M{\"u}ller, N.~Kjaergaard, and E.~S. Polzik,
  \enquote{Nondestructive probing of rabi oscillations on the cesium clock
  transition near the standard quantum limit,} Phys. Rev. Lett. \textbf{100},
  103601 (2008).

\bibitem{Kuzmich:1999kl}
A.~Kuzmich, L.~Mandel, J.~Janis, Y.~E. Young, R.~Ejnisman, and N.~P. Bigelow,
  \enquote{Quantum nondemolition measurements of collective atomic spin,} Phys.
  Rev. A \textbf{60}, 2346--2350 (1999).

\bibitem{Oblak:2005nx}
D.~Oblak, P.~G. Petrov, C.~L. Garrido~Alzar, W.~Tittel, A.~K. Vershovski, J.~K.
  Mikkelsen, J.~L. S\o{}rensen, and E.~S. Polzik,
  \enquote{Quantum-noise-limited interferometric measurement of atomic noise:
  Towards spin squeezing on the cs clock transition,} Phys. Rev. A \textbf{71},
  043807 (2005).

\bibitem{Appel:2009oq}
J.~Appel, P.~J. Windpassinger, D.~Oblak, U.~B. Hoff, N.~Kjaergaard, and E.~S.
  Polzik, \enquote{Mesoscopic atomic entanglement for precision measurements
  beyond the standard quantum limit,} Proceedings of the National Academy of
  Sciences \textbf{106}, 10960--10965 (2009).

\bibitem{Shiga:2012uq}
N.~Shiga and M.~Takeuchi, \enquote{Locking the local oscillator phase to the
  atomic phase via weak measurement,} New Journal of Physics \textbf{14},
  023034 (2012).

\bibitem{Chan:1983de}
H.P.~Yuen and V.W.S.~Chan, \enquote{{Noise in homodyne and heterodyne detection},}
  Optics Letters \textbf{8}, 177--179 (1983).

\bibitem{Shapiro:1984jp}
J.~Shapiro and S.~Wagner, \enquote{{Phase and amplitude uncertainties in
  heterodyne detection},}IEEE Journal of  Quantum Electronics \textbf{20},
  803--813 (1984).

\bibitem{Shapiro:1985gd}
J.~Shapiro, \enquote{{Quantum noise and excess noise in optical homodyne and
  heterodyne receivers},}  IEEE Journal of Quantum Electronics \textbf{21},
  237--250 (1985).

\bibitem{Leong:1986wb}
K.~Leong and J.~Shapiro, \enquote{{Phase and amplitude uncertainties in
  multimode heterodyning},} Optics Communications \textbf{58}, 73--77 (1986).

\bibitem{Haus:1995el}
H.~A. Haus, \enquote{{From classical to quantum noise},} J. Opt. Soc. Am. B \textbf{12},
  2019--2036 (1995).

\bibitem{Dariano:1994hu}
G.~D~ariano and M.~Paris, \enquote{{Lower bounds on phase sensitivity in ideal
  and feasible measurements},} Phys. Rev. A \textbf{49}, 3022--3036
  (1994).

\bibitem{Ou:1996jg}
Z.~Ou, \enquote{{Complementarity and Fundamental Limit in Precision Phase
  Measurement},} Phys. Rev. Lett. \textbf{77}, 2352--2355 (1996).

\bibitem{Luis:1996bb}
A.~Luis and J.~Pe{\v r}ina, \enquote{{Optimum phase-shift estimation and the
  quantum description of the phase difference},} Phys. Rev. A \textbf{54},
  4564--4570 (1996).

\bibitem{Ou:1997ks}
Z.~Y. Ou, \enquote{{Fundamental quantum limit in precision phase measurement},}
  Phys. Rev. A \textbf{55}, 2598--2609 (1997).

\bibitem{Opatrny:1998iu}
T.~Opatrn{\'y}, M.~Dakna, and D.-G. Welsch, \enquote{{Number-phase uncertainty
  relations: Verification by balanced homodyne measurement},} Phys. Rev. A
  \textbf{57}, 2129--2133 (1998).

\bibitem{Lye:2003vl}
J.~Lye, J.~Hope, and J.~Close, \enquote{{Nondestructive dynamic detectors for
  Bose-Einstein condensates},} Phys. Rev. A \textbf{67}, 043609 (2003).

\bibitem{Smith:2003ue}
G.~Smith, S.~Chaudhury, and P.~Jessen, \enquote{{Faraday spectroscopy in an
  optical lattice: a continuous probe of atom dynamics},} J. Opt. Soc. Am. B \textbf{5}, 323--329 (2003).

\bibitem{Lye:2004ys}
J.~Lye, J.~Hope, and J.~Close, \enquote{Rapid real-time detection of cold atoms
  with minimal destruction,} Phys. Rev. A \textbf{69}, 023601 (2004).

\bibitem{Kohnen:2011}
M.~Kohnen, P.~G. Petrov, R.~A. Nyman, and E.~A. Hinds, \enquote{Minimally
  destructive detection of magnetically trapped atoms using
  frequency-synthesized light,} New Journal of Physics \textbf{13}, 085006
  (2011).

\bibitem{Figl:2006}
C.~Figl, L.~Longchambon, M.~Jeppesen, M.~Kruger, H.~A. Bachor, N.~P. Robins,
  and J.~D. Close, \enquote{Demonstration and characterization of a detector
  for minimally destructive detection of bose condensed atoms in real time,}
  Applied Optics \textbf{45}, 3416 (2006).

\bibitem{Metcalf:1999}
H.~J. Metcalf and P.~van~der Straten, \emph{Laser Cooling and Trapping}
  (Springer, 1999).

\bibitem{Durfee:2006li}
D.~S. Durfee, Y.~K. Shaham, and M.~A. Kasevich, \enquote{Long-term stability of
  an area-reversible atom-interferometer sagnac gyroscope,} Phys. Rev. Lett.
  \textbf{97}, 240801 (2006).

\bibitem{Grimm:2000tp}
R.~Grimm, M.~Weidem{\"u}ller, and Y.~B. Ovchinnikov, \enquote{Optical dipole
  traps for neutral atoms,}  (Academic Press, 2000), pp. 95 -- 170.

\bibitem{Born:1999ws}
M.~Born and E.~Wolf, \emph{Principles of Optics} (Cambridge University Press,
  1999).

\end{thebibliography}
\end{document}